\documentclass{llncs}
\usepackage{epsfig}
\usepackage{makeidx}  
\usepackage{url}
\begin{document}
\title{Experimental DML over digital repositories in Japan}
\author{Takao Namiki,Hiraku Kuroda and Shunsuke Naruse}
\authorrunning{T. Namiki, H. Kuroda and S. Naruse}
\institute{Department of Mathematics, Hokkaido University,
  060-0810 Sapporo, Japan}
\maketitle

\begin{abstract}
In this paper the authors show an overview of Virtual Digital
Mathematics Library in Japan (DML-JP), contents of which consist of
metadata harvested from institutional
repositories in Japan and digital repositories in the world.
DML-JP is, in a sense, a subject specific repository which 
collaborate with various digital repositories.
Beyond portal website, DML-JP provides subject-specific metadata
through OAI-ORE.  By the schema it is enabled that digital
repositories can load the rich metadata which were added by
mathematicians. 
\end{abstract}

\section{Introduction and Backgrounds}
\label{intro}
In Japan about 70000
mathematical articles which had been reviewd in Math. Reviews were
published in 400 journal titles \cite{dml08}.  Nowadays electronic
edition of these journal titles are loaded on various digital
repositories, which are partly supported by SPARC Japan \cite{sj} 
and CSI project \cite{csi,csi-rep}.  Among such digital repositories one major
repository is projecteuclid.org and the other is institutional
repositories in Japan. 

Contributions
of these articles for DML for journal titles published in Japan are so 
important that we were planning to establish the potal website of the
articles.  Until Nov. 2008 about 20 small scale mathematical journals were
loaded on institutional repositories \cite{dml08,or08} and since 2005
projecteuclid.org have loaded 10 major mathematical journals published
in Japan.  

Considering these backgrounds, we constructed an experimental DML-JP
as a portal website based on metadata harvesting. The titles joined
with DML-JP are shown in the following.

\def\baselinestretch{0.8}
\begin{verbatim}
  Bull. Tokyo Gakugei University Sec. I
  Bulletin of College of Science the University Ryukyu
  Hiroshima Math. J.
  Hokkaido Mathematical Journal
  J. Math. Soc. Japan
  Japan J. Indust. Appl. Math.
  Journal of Mathematical Sciences, The University of Tokyo
  Journal of the Faculty of Education, Kagoshima University
  Journal of the Faculty of Science Shinshu University
  Journal of the Faculty of Science, Kagoshima University
  Journal of the Faculty of Science, the University of Tokyo
    Sect 1 A
  Journal of the Faculty of Science, Yamagata University
  Nagoya Math. J.
  Kodai Math. J.
  Nat. Sci. J. Fac. Educ. Hum. Sci. Yokohama National University
    Sec. I
  Natur. Sci. Report. Ochanomizu. Univ.
  Nihonkai Mathematical Journal
  Osaka J. Math.
  Proc. Japan Acad. Ser. A Math. Sci.
  Publ. Res. Inst. Math. Sci.
  Reports of the Faculty of Science and Engineering,
    Saga University. Mathematics
  RIMS Kokyuroku
  Ryukyu Mathematical Journal
  Sci. Rep. Yokohama National University Sec. I
  The science reports of the Kanazawa University
  Tohoku Math. J.
  Tokyo J. of Math.
  Tsukuba Journal of Mathematics
\end{verbatim}
\def\baselinestretch{1.0}

\section{Implementation}
Platform of DML-JP is based on EPrints 3.1.1 software. We choosed the
software because it was widely used and actively developed.  As
described above main contents of DML-JP are metadata harvested from
digital repositories and the first work for DML-JP is to transform the
harvested metadata into the format which is suitable for the platform.
In this section we show the part.

\subsection{Metadata harvesting}
A standard metadata format for institutional repositories in Japan is
junii2 format which is suitable to describe bibliographic information
of journal articles.  For projecteuclid.org and arxiv.org we choosed
oai\_dc format because these repository provides integrated
bibliographic information in dc:identifier elements.

\begin{figure}[hbtp]
  \centering
  \includegraphics{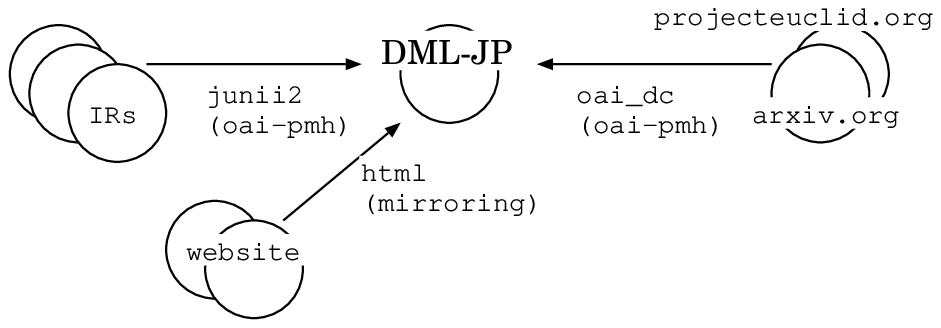}
  \caption{Metadata harvesting}
  \label{fig:m}
\end{figure}

Because our target is mathematical journals published in Japan, as
shown in \cite{dml08}, we are harvesting 16 institutional repositories
and two subject repositories.  The number of articles is over 30000.
This result means that about the half of all articles published in
Japanese mathematical journals are grasped.

We should manage two metadata formats.  The one is oai\_dc which project
euclid provides.  The other is junii2 format which is standard format
for institutional repositories for metadata exchange in Japan.
The following is an example of oai\_dc format metadata which is
provided from Project Euclid.

\def\baselinestretch{0.8}
\begin{verbatim}
<record>
<header>
<identifier>
  oai:CULeuclid:euclid.jmsj/1240435759
</identifier>
<datestamp>2009-04-23</datestamp>
<setSpec>jmsj</setSpec>
</header>
<metadata>
<oai_dc>
  <dc:title>
   Minimal 2-regular digraphs with given girth
  </dc:title>
  <dc:creator>BEHZAD, Mehdi</dc:creator>
  <dc:subject>05C20</dc:subject>
  <dc:publisher>Mathematical Society of Japan</dc:publisher>
  <dc:date>1973-01</dc:date>
  <dc:type>Text</dc:type>
  <dc:format>application/pdf</dc:format>
  <dc:identifier>
   http://projecteuclid.org/euclid.jmsj/1240435759
  </dc:identifier>
  <dc:identifier>
   J. Math. Soc. Japan 25, no. 1 (1973), 1-6
  </dc:identifier>
  <dc:identifier>doi:10.2969/jmsj/02510001</dc:identifier>
  <dc:language>en</dc:language>
  <dc:rights>
   Copyright 1973 Mathematical Society of Japan
  </dc:rights>
</oai_dc:dc>
</metadata>
</record>
\end{verbatim}
\def\baselinestretch{1.0}

One of the difficulty of the oai\_dc format above is
analysis of bibliographic information in {\tt dc:identifier} element
owing to there exist various journal name, series, volume and issue
format, which limitations of oai\_dc specification involve.

The following is an example of junii2 format metadata which is
provided from 
an institutional repository in Japan. 

\def\baselinestretch{0.8}
\begin{verbatim}
<record>
<header>
<identifier>oai:teapot.lib.ocha.ac.jp:10083/843</identifier>
<datestamp>2007-07-02T06:30:00Z</datestamp>
<setSpec>hdl_10083_792</setSpec>
</header>
<metadata>
<meta xmlns="http://ju.nii.ac.jp/junii2"
  xmlns:xsi="http://www.w3.org/2001/XMLSchema-instance"
  xsi:schemaLocation="http://ju.nii.ac.jp/junii2
  http://www.nii.ac.jp/irp/info/junii2.xsd">
<title>
 CONDITIONALLY TRIMMED SUMS FOR INDEPENDENT RANDOM VARIABLES
</title>
<creator>KASAHARA, Yuji</creator>
<NDC>400</NDC>
<publisher>Ochanomizu University</publisher>
<type>Article</type>
<NIItype>Departmental Bulletin Paper</NIItype>
<format>application/pdf</format>
<format>191755 bytes</format>
<URI>http://hdl.handle.net/10083/843</URI>
<fullTextURL>
http://teapot.lib.ocha.ac.jp/ocha/bitstream/10083/843/1/KJ00004470846.pdf
</fullTextURL>
<issn>00298190</issn>
<NCID>AN00033958</NCID>
<jtitle>Natur. Sci. Rep. Ochanomizu Univ.</jtitle>
<volume>46</volume>
<issue>2</issue>
<spage>9</spage>
<epage>12</epage>
<dateofissued>1995-12-30</dateofissued>
</meta>
</metadata></record>
\end{verbatim}
\def\baselinestretch{1.0}

An advantage of the junii2 format above is that each
bibliographic element is defined as an entity, which makes it easy to
retrieve bibliographic information, however, some institutional
repository does not include journal title in English and even if
included the expression does not coincide the expression of
Math. Reviews. By that reason it is relatively hard to retrieve MR
code and MSC from Math. Reviews database. 
Moreover,  Japanese letters are included
within several fields in original metadata. 

The two metadata formats were transformed into EPrints XML
format. Once bibliograhic information is retrieved, it is easy.
The following is an example of EPrints XML format. For DML-JP field
msc\_p, msc and mr were added to default configuration.

\def\baselinestretch{0.8}
\begin{verbatim}
<?xml version="1.0" encoding="utf-8" ?>
<eprints>
  <eprint xmlns="http://eprints.org/ep2/data/2.0">
  <rev_number>1</rev_number>
  <eprint_status>archive</eprint_status>
  <userid>1</userid>
  <metadata_visibility>show</metadata_visibility>
  <type>article</type>
  <ispublished>pub</ispublished>
  <subjects>
    <item>20-xx</item><item>QA</item>
  </subjects>
  <refereed>TRUE</refereed>
  <full_text_status>public</full_text_status>
  <date_type>published</date_type>
  <publication>Natur. Sci. Report. Ochanomizu. Univ.</publication>
  <datestamp>2007-08-01T01:50:05Z</datestamp>
  <title>
  Note on the Schur multiplier of a certain semidirect product
  </title>;
  <creators_name><item><family>Horie</family>
    <given>Mitsuko</given></item></creators_name>
  <official_url>http://hdl.handle.net/10083/839</official_url>
  <pagerange>85-88</pagerange>
  <volume>45</volume>
  <date>1994-12-15</date>
  <publisher>Ochanomizu Univeristy</publisher>
  <msc_p>20J06</msc_p>
  <msc><item>20C25</item></msc>
  <mr>1317509</mr>
  <related_url><item>
   <url>http://www.ams.org/mathscinet-getitem?mr=1317509</url>
   <type>MathSciNet</type></item></related_url>
  </eprint>
</eprints>
\end{verbatim}
\def\baselinestretch{1.0}

\subsection{Metadata managements}
From viewpoint of mathematical communication, there are several
metadata entries for an article which describe mathematical
classifications, reviews and locations of preprints.  An identifier
of an article is MR number which specify the review in Math. Reviews
published by American Mathematical Society (AMS) in
the form
\url{http://www.ams.org/mathscinet-getitem?mr=}{\it{}id\_number}.
AMS also provides Mathematics Subject Classification which is a
comprehensive classification for mathematical literatures. 
Authors of mathematical literatures are required to specify at least one
classification.
In mathematics and theoretical physics preprints play an important
role for scholarly communication.  It is necessary for researchers to
know the locations of preprints for each article which have not been
published in any journals.

Considering the resercher's needs, the set of harvested metadata is
not necessarily enough to describe each article.

\begin{figure}[hbtp]
  \centering
  \includegraphics{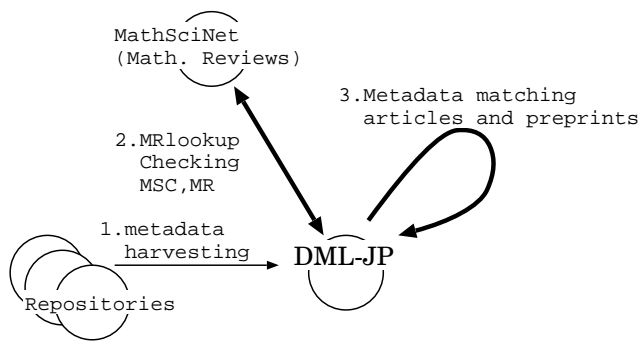}
  \caption{Metadata processing}
  \label{fig:m1}
\end{figure}

\subsection{Examples}
A typical example is an entry
\url{dmljp.math.sci.hokudai.ac.jp/32786/}.  From the url we can get
the information in the following table.
Prefix IR means that the entry was retrieved from IR and MR means
Math. Reviews.

\begin{verbatim}
  IR Author: Maeda, Masao
  IR Title: The four-or-more Vertex Theorems in 2-dimensional
     Space Forms
  IR Citation: Nat. Sci. J. Fac. Educ. Hum. Sci. Yokohama National
               University Sec. I, 1 (1998) . pp. 43-46.
  IR Official URL: http://hdl.handle.net/10131/1069
  MR MSC Primary:   53A35, 53A, 53
  MR MSC Secondary: 53A04, 53A, 53
  MR Math. Reviews ID: 1710269
  MR Review URL: http://www.ams.org/mathscinet-getitem?mr=1710269
\end{verbatim}

Though this journal is so small and interdisciplinary
that only this article is reviewed and indexed in Math. Reviews, you
can find in the review URL that this article was cited from a review
article in the field.

\subsection{Preparation for OAI-ORE and SWORD}
Owing to full text PDF files of DML-JP are stored in digital
repositories and the quality of their mathematical metadata is not
enough for institutional repositories,
MR code and MSC should be reflected to add the value for the
repository contents. For these type of collaboration, we would like to
provide the metadata by OAI-ORE Atom Serialization format.

In the Resource Map of 
each article we aggregate official url and DML-JP url.  In the entry
of DML-JP we prepare an XML file in METS metadata format which could
be imported to original repositories via SWORD protocol which 
many digital repository communities have already choosed as inter
repository interfaces.

We intend to establish resource finding and exchange schema between
digital repositories by the implementation, which is merely
experimental phase.

\begin{figure}[hbtp]
  \centering
  \includegraphics{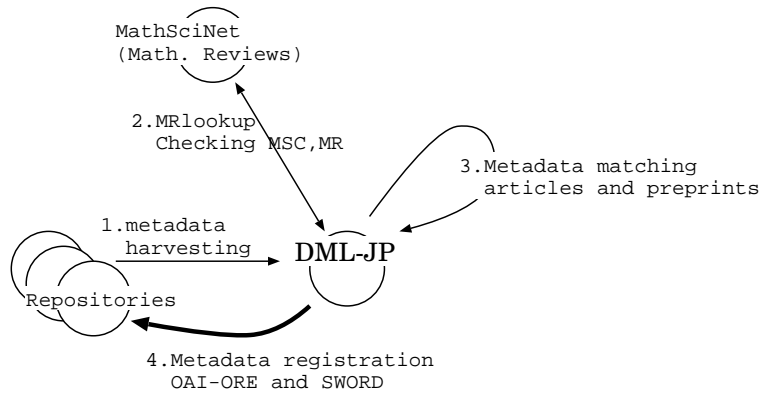}
  \caption{Metadata registration}
  \label{fig:ore}
\end{figure}

Though only OAI-ORE Atom serialization for official url was
implemented at this time, 
METS format metadata is easily generated by a function of EPrints.
So the implementation of the picture above will be realized within a
year.  The following is a part of an example of ORE Atom
serialization.

\begin{verbatim}
 <!-- Aggregated Resources -->
 <atom:link href='http://projecteuclid.org/euclid.kmj/1138846413'
   title='A remark on derived spaces'
   rel='http://www.openarchives.org/ore/terms/aggregates' />
 <atom:link href='http://projecteuclid.org/euclid.tmj/1192117987'
   title='Spectral synthesis in the Fourier algebra and the
      Varopoulos algebra'
   rel='http://www.openarchives.org/ore/terms/aggregates' />
\end{verbatim}

\section{Statistics in DML-JP}\label{stat}
In this section we show statistics of the journals which are the
targets of DML-JP.  The first result is performance of the journals
published in Japan.

\subsection{The number of articles for each research fields}
The first result is the percentage in the journal articles and
whole articles within each MSC shown in Table \ref{tab:statistics},
which is retrieved from Math. Reviews.  There are 12
research fields in which the table shows the article share from 5
percent to 10 percent. These research fields also are active in Japan
from mathematician's intuition. DML-JP covers about 50 \% of the
articles and moreover there are several mathematical papers which are
not indexed by Math. Reviews.

\begin{table}
  \centering
  \begin{tabular}{r|c|l}\hline
  \% & Articles/Total & MSC Primary\\\hline\hline
  10.62 & (1923/18103) & 57 Manifolds and cell complexes\\
  10.00 & (1852/18506) & 32 Several complex variables and analytic spaces\\
  9.48 & (545/5748) & 31 Potential theory \\
  9.46 & (1048/11077) & 55 Algebraic topology\\
  9.20 & (1902/20655) & 14 Algebraic geometry \\
  8.15 & (3307/40538) & 53 Differential geometry\\
  7.68 & (875/11392) & 13 Commutative rings and algebras\\
  7.45 & (525/7041) & 12 Field theory and polynomials\\
  6.58 & (2301/34968) & 11 Number theory\\
  6.25 & (734/11742) & 22 Topological groups, Lie groups\\
  5.84 & (1922/32891) & 30 Functions of a complex variable\\
  5.44 & (1305/23970) & 16 Associative rings and algebras\\\hline
  \end{tabular}
  \caption{Performance of the journal}
  \label{tab:statistics}
\end{table}

\subsection{Application of HITS algorithm}
The second result is based on HITS \cite{Ferahat,Lampel} algorithm
which is widely applied in ranking of webpages.  Because HITS itself is
for weighted directed graph we can apply it for relation of research
fields if we make such a graph for them. For structure matrix $M$ of a
weighted directed graph hub score (H-score) of a node is defined as
the value of the correspondent element of maximal eigenfunction for
$M^tM$ and authority score (A-score) as the value of $MM^t$.

\begin{figure}
  \centering
  \includegraphics{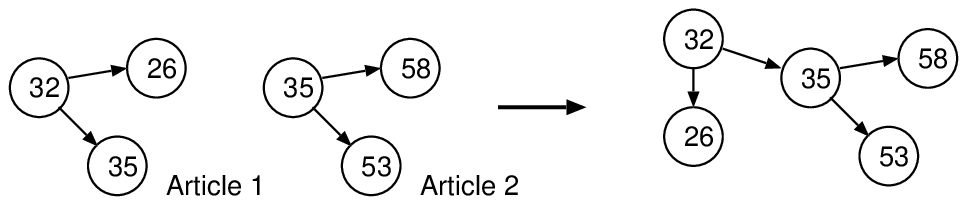}
  \caption{Fragment of directed graph generated by two article.}
  \label{fig:g1}
\end{figure}

Let the node set be the first two digits of MSC.  If an article
specify MSC Primary A and MSC Secondary B, we set an edge from A to B
and add weight 1 to the edge, which is a fragment of the graph
constructed from an article. We have whole graph for all of target
articles as a result. Figure \ref{fig:g1} shows the process and the
directed graph for each year is shown in URL \url{oaia.math.sci.hokudai.ac.jp/navi/jp-a/}.

\begin{figure}
  \centering
  \begin{tabular}{cc}
  \includegraphics[scale=.4]{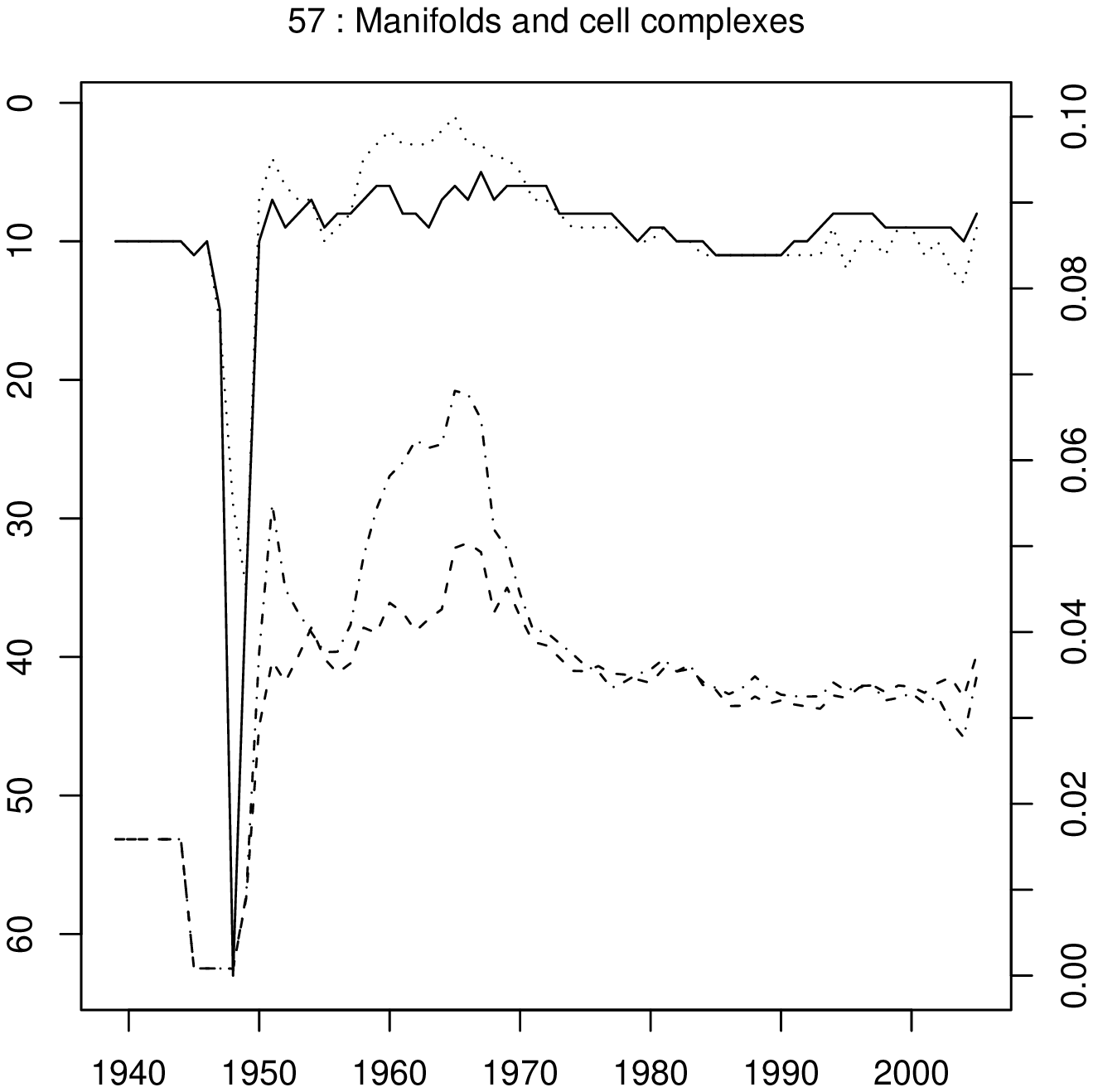}&
  \includegraphics[scale=.4]{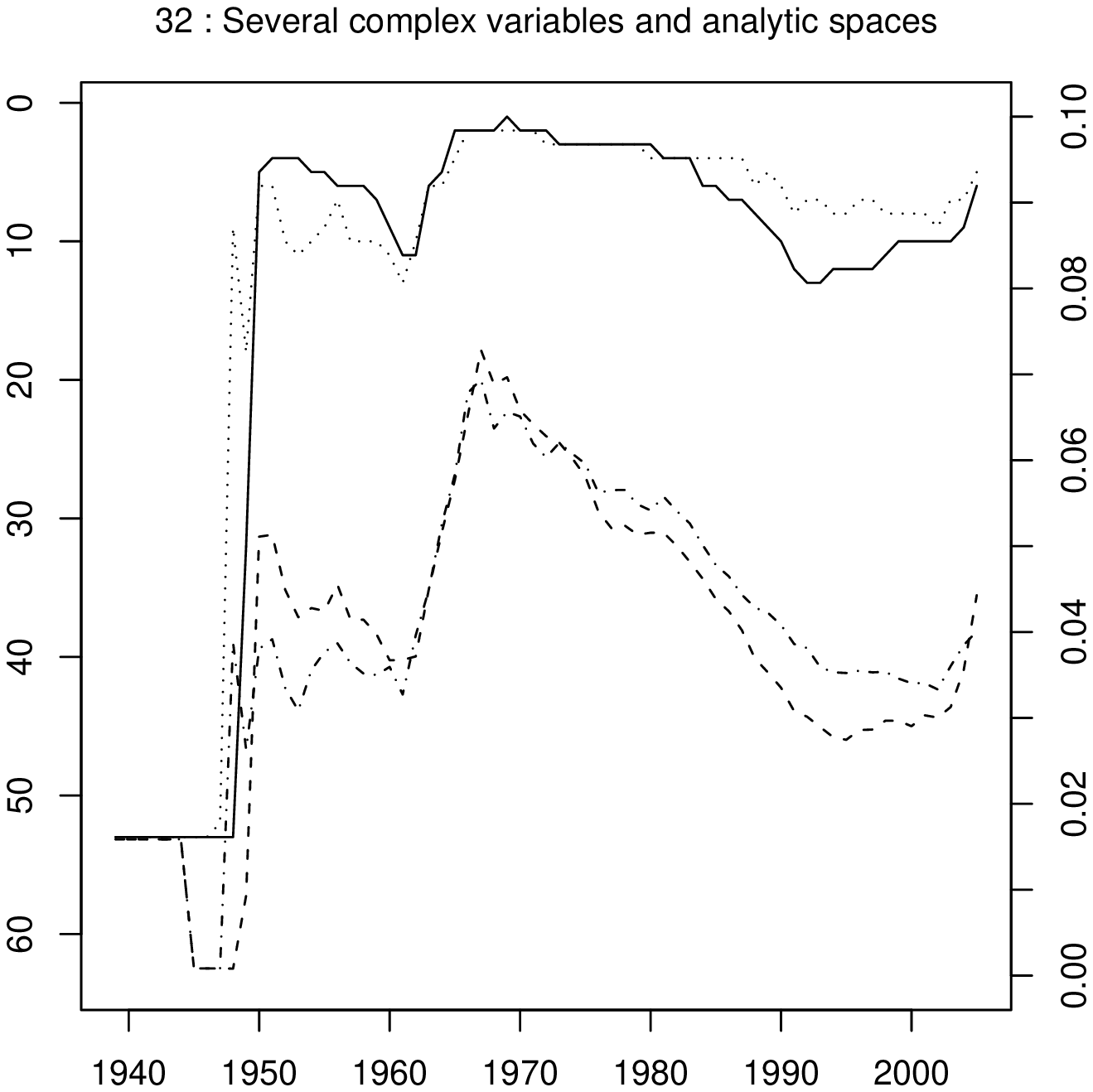}\\
  \includegraphics[scale=.4]{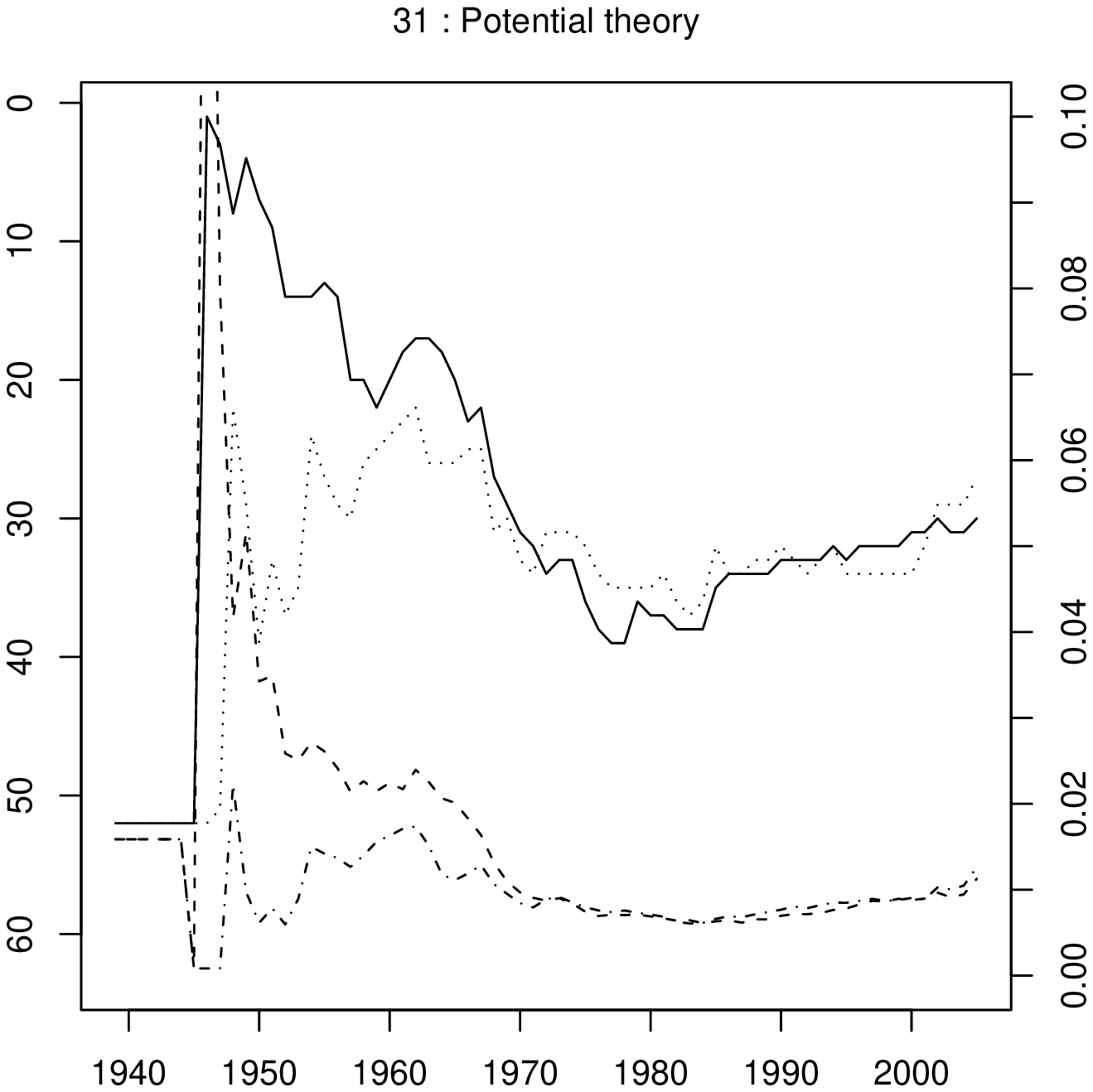}&
  \includegraphics[scale=.4]{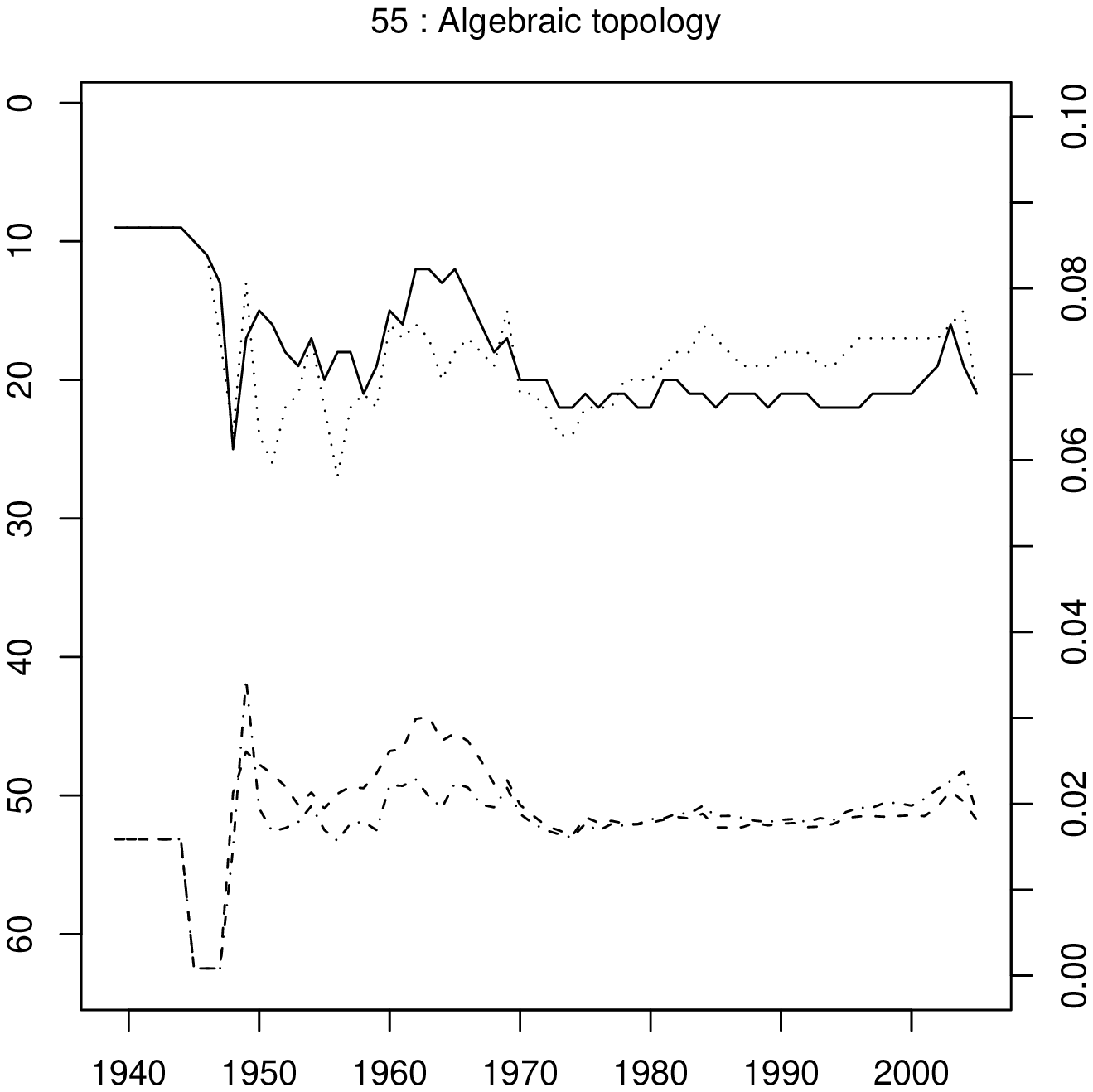}\\
  \includegraphics[scale=.4]{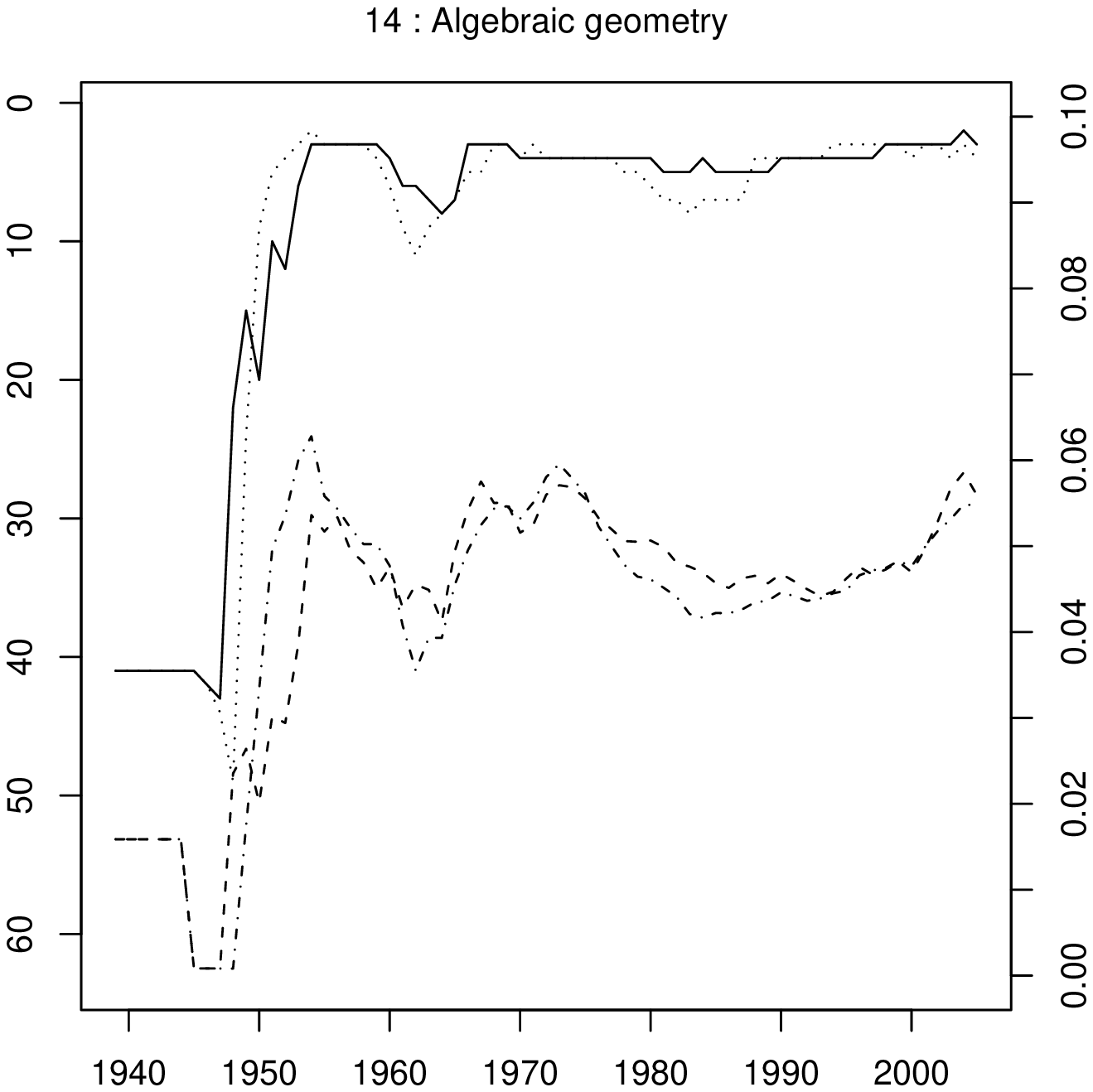}&
  \includegraphics[scale=.4]{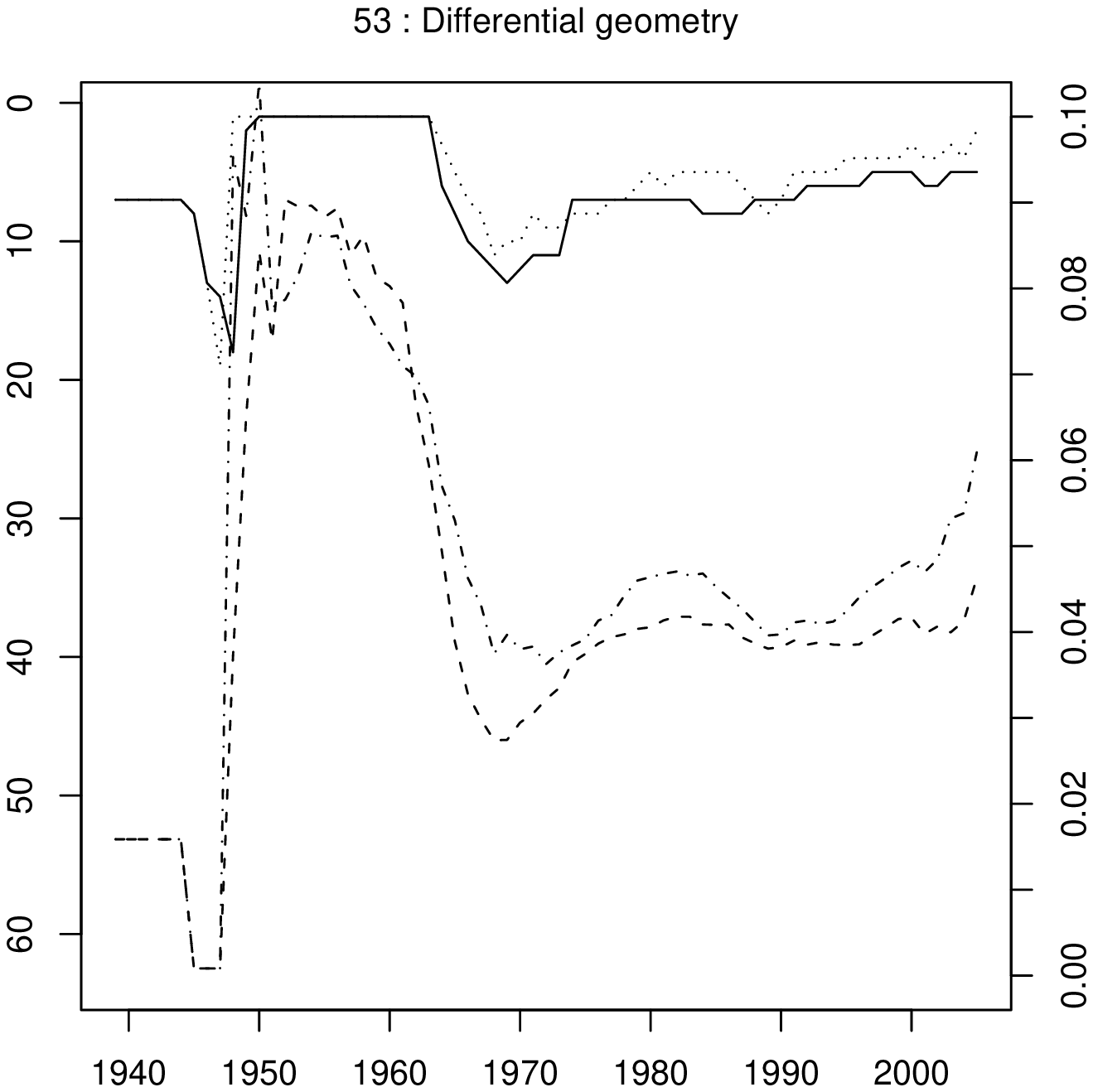}
  \end{tabular}
  \caption{Time series of H-score, A-score and ranking by the
    scores. Solid line: H-score, Dotted line: A-score, Broken line:
    value of H-score, Dotted broken line: value of H-score.}
  \label{fig:H}
\end{figure}

Figures \ref{fig:H} shows time
series of ranking by H-score and A-score and the scores themselves for
the first six research fields in Table \ref{tab:statistics}. 
The value for each
year of the time series is calculated by the articles published in the
following ten years. For example the value of 1990 is from the graph
generated by articles published from 1990 to 1999.

\begin{figure}
  \centering
  \begin{tabular}{cc}
  \includegraphics[scale=.4]{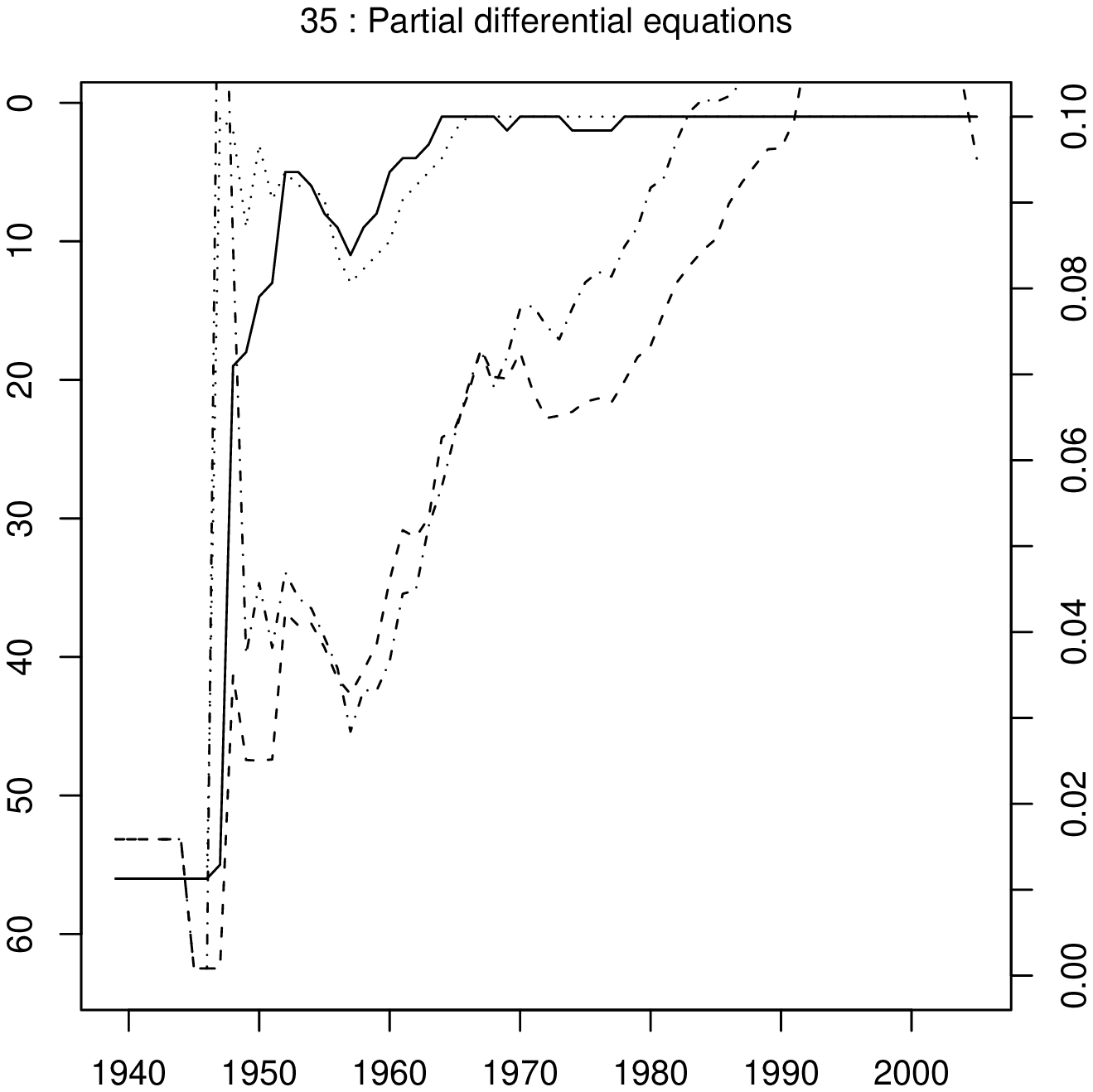}&
  \includegraphics[scale=.4]{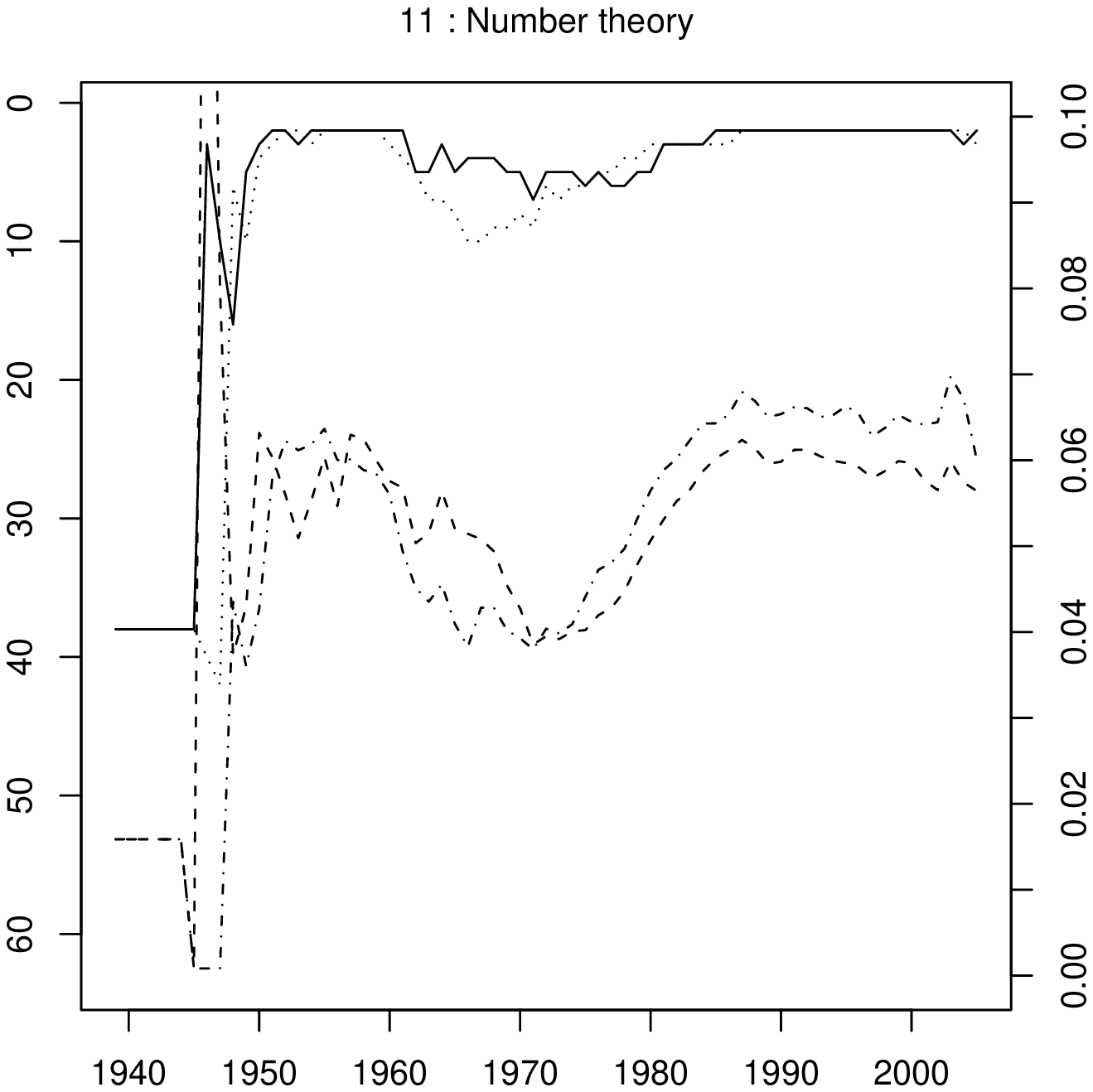}
  \end{tabular}
  \caption{HITS results for MSC 35 and 11}
  \label{fig:35-11}
\end{figure}

We can see that the ranking from these scores does not coincide the
ranking of Table \ref{tab:statistics} and moreover is not in
proportion to the number of articles.
On one hand,
in the six research fields shown in Figure \ref{fig:g1}, tendency of
the scores and ranks of Potential theory (MSC 31) and Algebraic
topology (MSC 55) is not so high as shown in Table
\ref{tab:statistics}.
On the other hand,
Figure \ref{fig:35-11} shows the typical difference.  In
Partial differential equation (MSC 35), there are huge number of
articles are published in the world.  So
the performance is estimated relatively low in Table
\ref{tab:statistics}.  Despite that, the scores and ranks shown in
Figure \ref{fig:35-11} are high.
In Number theory, it means that the influence is strong more than the
ranking of the number of published articles.

\section{Discussions and Future work}
DML-JP is a collaborating work with librarians and mathematical
community.  From a view of applied mathematics, it is essentially
important to disseminate these articles on subject portal website for
widely usable objectives.  In Section \ref{stat} we show several
viewpoint to represent activities on mathematics in Japan.  It is
important to have various methods to estimate the activities.

Unfortunately OCR technology for mathematical expression is not
familliar with community outside mathematical publishing.  We are
planning to mirroring OA articles of these journals and providing
full-text by xml+mathml format with certain presision as far as
possible. 

Though DML-JP introduced in the article is experimental DML, we
consider that we can develope it by the advantage of metadata based
repository.

\section{Acknowledgement}
This work is mainly supported by SPARC JAPAN \cite{sj} from Apr. 2008 
to Mar. 2009, and had been supported  by Department of mathematics,
Hokkaido University under governmental funding of 21st century Center
of Excellence. Mathematical Society of Japan also supports this
activity.

\end{document}